\documentclass{ws-ijmpd}
\usepackage[super,compress]{cite}
\usepackage{url}

\usepackage[dvips,breaklinks]{hyperref}
\hypersetup{colorlinks,urlcolor=black,citecolor=black,linkcolor=black,filecolor=black}
\usepackage{breakurl}
\def\vr{{\bf r}}
\def\l{\lambda}
\begin{document}

\markboth{Kroupa, Pawlowski \& Milgrom}
{The failures of the standard model of cosmology require a new paradigm}

%
\catchline{}{}{}{}{}
%

\title{
The failures of the standard model of cosmology require a new paradigm
}

\author{Pavel Kroupa$^*$ and Marcel Pawlowski$^{**}$
}

\address{Argelander-Institute for Astronomy, University of Bonn,\\
Auf dem H\"ugel 71, D-53121 Bonn, Germany\\
$^*$: pavel@astro.uni-bonn.de\\
$^{**}$: mpawlow@astro.uni-bonn.de
}

\author{Mordehai Milgrom}

\address{Department of Particle Physics and Astrophysics\\
Weizmann Institute of Science\\
Rehovot 76100, Israel\\
moti.milgrom@weizmann.ac.il
}

\maketitle

\begin{history}
\received{Day Month Year}
\revised{Day Month Year}
\end{history}

\begin{abstract}
  Cosmological models that invoke warm or cold dark matter can not
  explain observed regularities in the properties of dwarf galaxies,
  their highly anisotropic spatial distributions, nor the correlation
  between observed mass discrepancies and acceleration. These problems
  with the standard model of cosmology have deep implications, in
  particular in combination with the observation that the data are
  excellently described by Modified Newtonian Dynamics (MOND). MOND is
  a classical dynamics theory which explains the mass discrepancies in
  galactic systems, and in the universe at large, without invoking
  `dark' entities. MOND introduces a new universal constant of nature
  with the dimensions of acceleration, $a_0$, such that the
  pre-MONDian dynamics is valid for accelerations $a \gg a_0$, and the
  deep MONDian regime is obtained for $a\ll a_0$, where space-time
  scale invariance is invoked.  Remaining challenges for MOND are (i)
  explaining fully the observed mass discrepancies in galaxy clusters,
  and (ii) the development of a relativistic theory of MOND that will
  satisfactorily account for cosmology. The universal constant $a_0$
  turns out to have an intriguing connection with cosmology: $
  \bar{a}_0 \equiv 2 \pi a_0 \approx c H_0 \approx
  c^2(\Lambda/3)^{1/2}$. This may point to a deep connection between
  cosmology and internal dynamics of local systems.
\end{abstract}

\keywords{cosmology; dark matter; gravitation; MOND}

\ccode{PACS numbers: 98.35.-a;  98.56.-p; 98.62.-g; 98.65.-r;
  98.80.-k;
04.20.Cv; 04.80.Cc; 95.35.+d; 98.80.Es
}


\section{Introduction}
\label{kpmsec:introd}

Newton\cite{Newton1760} formulated his dynamics subject to the
observational constraints from the Solar system, while Einstein
developed his general relativistic field equation\cite{Einstein16}
with the prerequisite that they reproduce Newton's work in the
classical, non-relativistic limit. General relativity has successfully
passed tests in the astronomically small-spatial-scale and strong
acceleration limit (Solar system and smaller in length scale and
stronger in acceleration scale).  The presently favoured understanding
of cosmological physics is based on assuming Einstein's field equation
to hold on galactic and larger scales and for very small accelerations
as are found in galactic systems (postulate 1). In addition, it is
assumed that all matter emerged in the Big Bang (postulate~2). The
first two postulates lead to an inhomogeneous and highly curved
cosmological model which is in disagreement with the observed
distribution of matter, unless inflation is additionally postulated to
drive the universe to near flatness and homogeneity briefly after the
Big Bang (postulate~3). The structures and their kinematics which
arise from this postulate again do not match the observed ones unless
cold or warm dark matter (DM) is postulated in addition to aid
gravitational clumping on galactic scales (postulate~4). This DM must,
as observations constrain, consist of non-relativistic particles which
only interact significantly through gravitation\cite{Blumenthal84}.
The so constructed model does not match the observed increase in the
rate of expansion unless it is additionally postulated that dark
energy drives an emerging new era of inflation (postulate~5). The
resulting five-postulate construction with its many parameters
determined using observational data\cite{Famaey12} is referred to as
the concordance cosmological model which is the currently generally
accepted cold or warm DM-based standard model of cosmology
(SMoC)\cite{FrenkWhite12, Silk12}.

The SMoC can be used to compute the distribution of matter on galactic
scales because it is based on Newtonian dynamics which is a linear and
thus, in principle, readily computable dynamics theory\cite{Kuhlen12,
  FrenkWhite12}. The complex processes of the baryons (heating,
cooling, star-formation, stellar and other feedback) are dealt with by
employing parametrised laws, but few rigorous convincing predictions
have come forth due to the tunability of the modelling, the lack of
adopted constraints from observed star formation processes, and due to
the haphazardness of the many mergers each DM halo
experiences\cite{Mutch12}. The statistical distribution of sub-halos
on the scales of many kpc around normal galaxies is among the rigorous
predictions of the SMoC.  The other rigorous prediction is that dwarf
galaxies formed from tidal material expelled during galaxy encounters
(see Sect. \ref{kpmsec:DDGT}) cannot contain substantial amounts of
DM.

The computations by many research groups have demonstrated that the
SMoC has significant problems, the number of which has been increasing
with improving computer power\cite{Kroupa12}.  While the greatest
problems (see below) are usually not discussed by DM advocates even
when they do discuss problems for the SMoC, e.g. in
Ref(\refcite{FrenkWhite12}), it is held by most dark-matter advocates
that virtually all problems mentioned by the respective authors can be
solved once modelling the baryonic processes is improved with larger
resolution\cite{Kuhlen12}. Thus, for example, one of the failures, the
missing satellite problem\cite{Klypin99,Moore99}, is deemed to have
been solved by now\cite{Brooks12}: according to the SMoC each DM halo
ought to contain a large number of satellite DM halos which are the
phase-space substructures that merge during the hierarchical formation
of larger structures\cite{Klypin11}. Milky Way and Andromeda class
galaxies ought to have many hundreds of satellite galaxies, each of
which are immersed in their own DM halos. The observed small number of
satellites (currently 24 are known for the Milky Way and a slightly
larger number is known for Andromeda) is then a result of the complex
and tunable baryonic processes cleaning out baryons from the vast
majority of satellite DM halos.  Some of the problems of the SMoC have
been used to investigate possible model extensions, for example by
invoking new dark forces between the DM particles which affect
structure growth on sub-kpc scales\cite{Cyr-Racine12}. These
additional postulates do not significantly affect the distribution of
sub-halos on scales of many kpc, as this is given by the hierarchical
infall of DM into larger structures.

The SMoC has been stated to be in agreement with the large-scale
distribution of matter (although the problems on the local 100~Mpc
and~16~Mpc scales undermine this statement,
Sec.~\ref{kpmsec:additionaltests}) and with the cosmic microwave
background (CMB) fluctuations\cite{FrenkWhite12}. But it can only
remain as a valid description of the real universe if every one of the
postulates individually pass observational tests.  In order to certify
the validity of the SMoC and its dark-sector variants {\it robustly}
against the observational data, tests must be developed which are
insensitive to the details of baryonic physics. Various such tests
have recently been devised\cite{Kroupa10}, and here the two most
important ones (Sec.~\ref{kpmsec:DDGT}, Sec.~\ref{kpmsec:phasespace})
with consistency checks (Sec.~\ref{kpmsec:additionaltests}) are
discussed. The observed correlation of kinematical properties of
galaxies over several orders of magnitude in acceleration
(Sec.~\ref{kpmsec:MDA}) convincingly demonstrates that a gravitational
theory which differs from Newton's is valid. The Modified Newtonian
Dynamics (MOND) theory\cite{Milgrom83} is in excellent agreement with
these data and provides an apparently correct description of dynamics
on galaxy scales (Sec~\ref{kpmsec:MOND}).

\section{The dual dwarf galaxy theorem}
\label{kpmsec:DDGT}

In any realistic cosmological theory two types of galaxies
emerge. Primordial (`type A') galaxies form directly after the Big
Bang from gravitational instabilities in the cooling baryonic
matter. Tidal dwarf galaxies (`type B') form from interacting,
rotationally-supported type A galaxies: from tidal arms, which
fragment and form dwarf galaxies as well as star clusters.  This is
evident in many observations of interacting
galaxies\cite{Duc94,Duc98,Bournaud10,Dabringhausen12} and in all
high-resolution simulations of interacting galaxies
\cite{Wetzstein07,Bournaud10,Fouquet12}.

Indeed, high-resolution simulations in the SMoC show that
type~A and type~B dwarfs have different dynamical and morphological
properties (see points~(i) and (ii) below). This comes about
because the formation of type~A dwarfs is dominated by the collapse
and mergers of dwarf DM halos in which the end products retain
  much of the DM. In the SMoC, a large number of these
sub-halos orbit as bound type~A sub-structures in the more massive
host halos,  {many more than are observed:} The missing
satellite problem therewith emerges (Sec.~\ref{kpmsec:introd}).
Type~B dwarfs, on the other hand, cannot capture significant amounts
of DM  {because they form from the `cold' parts of (disc)
  galaxies, which exclude the DM}.  In the SMoC so many tidal dwarf
galaxies are generated that  {to account for these,
  statistically,} most of  {the observed} dwarf elliptical (dE)
galaxies would  {have to} be type~B dwarfs\cite{Okazaki00}.

The SMoC thus predicts that two types of galaxies ought to be abundant
that have different dynamics: one showing DM, the other not. But in
fact of observation, rotating late-type dwarf galaxies as well as
faint dwarf-spheroidal galaxies show evidence for DM. A tidal dwarf
galaxy which is observed to have a DM component does not immediately
disprove the SMoC. DM may appear to be present in a tidal dwarf galaxy
if it is wrongly assumed to be in dynamical equilibrium when, in fact,
it is not, as it is being perturbed by a time varying tidal field from
the host galaxy\cite{Kroupa97,Casas12}.  And, a young tidal dwarf
galaxy may be accreting gas and may thus not be in rotational
equilibrium feigning a DM content.

All type~A galaxies which are rotationally supported are known to
precisely lie on the Baryonic-Tully-Fisher Relation
(BTFR)\cite{Famaey12} which therefore must be defined by the DM halo
if the SMoC is true. The physics responsible for the precise
conspiracy of the putative DM halo and galaxy luminosity to produce
the BTFR over many orders of magnitude in galaxy luminosity however
remains unknown.  Tidal dwarf galaxies that show evidence for DM
because they are out of equilibrium ought not to lie on the BTFR,
because the tidal perturbation of, or gas accretion onto, the dwarf
would not likely conspire to place a DM-free type~B galaxy onto the
BTFR of the type~A dwarfs.

The prediction made by the SMoC that there are two distinct classes of
dwarf galaxy (the 'dual dwarf galaxy' theorem) is falsified by
observations, as follows\cite{Kroupa12}:

(i) By invoking the BTFR of rotationally supported
galaxies\cite{Gentile07, mil07}: Type B dwarfs do not contain DM and
therefore cannot lie on the same BTFR of type~A galaxies in the SMoC.
But observations show them to lie on the same BTFR as type~A
galaxies. The rotational velocities of the type~A galaxies therefore
cannot be given by a DM halo.  Therewith, galactic DM halos cannot
exist and one of the major pillars of the SMoC (postulate~4) breaks
away such that the whole model is falsified.

(ii) By considering the radius--mass relation of pressure-supported
dwarf galaxies: dE galaxies are thought to be of type~A and thus to
form in putative DM halos. The final structural parameters of dE
galaxies and type B dwarfs therefore cannot agree. But observations
show type B dwarfs to be indistinguishable from dE
galaxies\cite{Dabringhausen12}. Thus, DM halos cannot play a role in
dwarf galaxies, leading to the same conclusion as under~(i) above.

\section{Significant anisotropic phase-space distribution of satellite
galaxies}
\label{kpmsec:phasespace}

The SMoC robustly predicts the primordial DM dominated (type~A)
satellite galaxies in a major halo to be distributed spheroidally
around the host galaxy. The infall of satellite galaxies from
filaments leads only to mild anisotropies\cite{Pawlowski12a}.

The Milky Way galaxy, being part of the universe, must conform to
these predictions. Instead the Galactic satellite system forms a vast
polar structure (VPOS) which is a disk-like
distribution\cite{Kroupa05} about 40~kpc thick and 400~kpc in diameter
containing all known satellite galaxies as well as all young halo
globular clusters and about half of all known stellar and gas
streams\cite{Pawlowski12b}.  The proper motion measurements of~10
satellite galaxies show~9 of the satellites to be orbiting within the
VPOS (one satellite is on a counter-rotating orbit in comparison to
the~8 others). One satellite, Sagittarius, orbits perpendicularly to
the VPOS and to the MW. To have~nine of ten type~A (DM sub-halo)
satellites with measured proper motions orbiting within the VPOS is
ruled out with very high confidence\cite{Libeskind09}.

Andromeda also has an anisotropic satellite system\cite{Metz07,Ibata12}, and a
few other galaxies are known which have satellite galaxies arranged in
highly correlated phase-space structures\cite{Kroupa12}.  The
prediction of the SMoC is therewith ruled out conclusively because the
Milky Way is not unique in this
property\cite{Kroupa10,Angus11a,Kroupa12}.

\section{Consistency tests}
\label{kpmsec:additionaltests}

The falsification\cite{Kroupa12} of the SMoC with the above argument
has deep implications. Can this  {argument} be erroneous?  A
number of auxiliary tests have been developed\cite{Kroupa10}, which
the SMoC again does not pass.  A visualisation of the results of the
comparison between many predictions and observational data is
presented in the theory confidence graph\cite{Kroupa12}.  If each of
the 22 listed failures were to be associated with a loss of confidence
of (only) 50~per cent that the SMoC accounts for a particular property
of the real universe, then according to the confidence graph, the
overall probability that the SMoC is a valid theory of the universe
would be $0.5$ raised to the power of $22$, corresponding to a
confidence of $2.3\times10^{-5}$~per cent.

Three examples of such problems are the cusp/core problem, the
downsizing and the missing-bright-satellites problem. These are
usually acknowledged as problems, but  {claimed to be (probably)
  solvable}\cite{FrenkWhite12,Brooks12,Kuhlen12,Bovill11,BK11}.  The
real problems for which no solution can be identified are usually not
mentioned in the DM literature: (a)~The dual dwarf galaxy theorem and
satellite anisotropy failures discussed above are catastrophic for the
SMoC. (b)~In the Local Volume with radius of 8~Mpc around the Sun, the
matter distribution in the Local Void and near the edges of the
filaments containing galaxies is incompatible with the expectations
from the SMoC\cite{Peebles10}.  (c)~The 50~Mpc radius region around us
is missing DM by a factor of~3--4 while the fluctuations in the
density must not be larger than about 10~per cent on these scales if
the SMoC were true\cite{Karachentsev12}. (d)~Disk galaxies of similar
mass are observed to show too little scatter in their
properties\cite{Disney08}, which implies that galaxies obey laws that
do not emerge in the SMoC.

\section{The mass-discrepancy-acceleration correlation: towards
  galaxy laws}
\label{kpmsec:MDA}

\begin{figure}[pb]
\centerline{\psfig{file=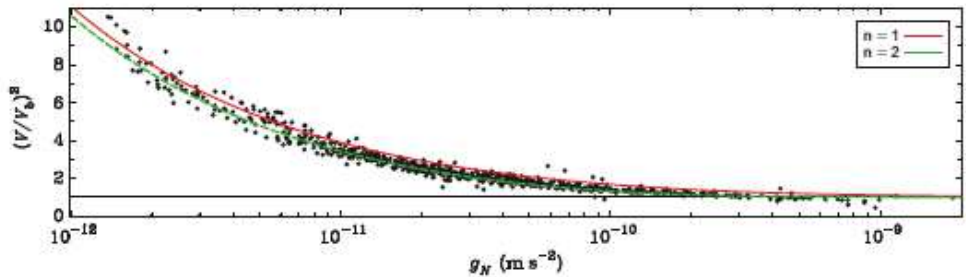,width=13cm}}
\vspace*{-3mm}
\caption{The mass-discrepancy--acceleration
  correlation\cite{McGaugh04,Famaey12} for disk galaxies. The
  red/solid and green/dashed curves are two different forms of the
  MONDian transition function as in fig.~11 in
  Kroupa\cite{Kroupa12}. For the Solar system $g_N>6\times
  10^{-5}$m/s$^2$. \label{kpmfig:mda}}
\end{figure}

The rotation velocity, $V(r)$, measured at a radius $r$ in a disk
galaxy can be compared to the rotation velocity, $V_b(r)$, that one
would expect due to only the observed baryonic matter. A plot of
$V^2/V_b^2$, which is the mass discrepancy at $r$, in a given galaxy
as a function of $r$, gives the putative DM profile of the galaxy. As
pointed out by McGaugh\cite{McGaugh04}, combining many such
measurements for many galaxies yields a mass-discrepancy vs $r$ plot
with no correlation. However, plotting the mass-discrepancy in
dependence of the baryonic-Newtonian acceleration, $g_N=V_b^2/r$,
yields a tight correlation (Fig.~\ref{kpmfig:mda}).
The mass-discrepancy--acceleration correlation deviates from the
Newtonian value below an acceleration of $a_0\approx
1.2\times10^{-10}$m/s$^2 = 3.8\,$pc/Myr$^2$ which is about five orders
of magnitude smaller than the acceleration of Neptune in the Solar
system.

No known physics of the DM particles can account for the observed
mass-discrepancy--acceleration correlation. But this relation is
convincing evidence for gravitational dynamics becoming non-Newtonian
at accelerations $a<a_0$. This and many other correlations were {\it
  predicted} 30~years ago by Milgrom\cite{Milgrom83} who introduced a
new constant of nature, $a_0$  {at the foundation of a new
  paradigm dubbed `Modified Newtonian Dynamics' (MOND). In this
  paradigm Newtonian dynamics and general relativity are modified, so
  that galaxy dynamics is explained without the need for dark matter.
  Since its formulation in 1983\cite{Milgrom83} MOND has passed all
  observational tests from $10^6\,M_\odot$ dwarf disk
  galaxies\cite{Famaey12} to massive elliptical galaxy\cite{Milgrom12}
  scales, with some tension remaining on the globular- and
  galaxy-cluster scales\cite{Kroupa12}. These are, however, not
  major\cite{Sanders07,Sanders12}.}

\section{The MOND paradigm}
\label{kpmsec:MOND}
While MOND has developed significantly in its particulars (such as
formulations of various underlying theories), its basic
non-relativistic (NR) tenets, from which follow all its major
predictions, remain the same as originally proposed. These tenets, put
in a somewhat improved form, are: (1) A new constant, $a_0$, with the
dimensions of acceleration is introduced into dynamics. (2) A MOND
theory must tend to standard dynamics in the limit $a_0\rightarrow 0$
(in other words, when the theory is applied to a system where all the
quantities of dimensions of acceleration, $a$, are much larger than
$a_0$). (3) In the opposite limit, $a_0\rightarrow\infty$ (namely,
when all $a\ll a_0$ in which case we also have to take $G\rightarrow
0$ so that $Ga_0$ remains fixed) the theory is postulated to become
space-time scale invariant\cite{milgrom09a}, namely, invariant under
$(t,\vr)\rightarrow\l(t,\vr)$.

Conceptually, $a_0$ thus plays a role as a demarcation acceleration
between the validity domain of the pre-MOND dynamics ($a\gg a_0$) and
the MOND regime ($a\lesssim a_0$). In the former, $a_0$ disappears
from physics, but in the latter domain $a_0$ appears with full impact
on various phenomena. These roles are similar to those of Planck's
constant in the classical--quantum context, and of the speed of light,
$c$, in the classical--relativistic context.  All these constants play
the role of demarcations between the old and new physics. Also, taking
them to the appropriate limit ($\hbar\rightarrow 0$ -- the
correspondence principle, and $c\rightarrow \infty$) takes one to the
old, classical theory, in which these constants do not appear. Also,
they all appear ubiquitously in the description of many, apparently
unrelated phenomena\footnote{For example, in quantum theory: the
  black-body spectrum, the photoelectric effect, the hydrogen
  spectrum, quantum Hall effect, etc.} in the new-physics regime. We
shall see below how $a_0$ enters in many such phenomena and effects in
low-acceleration galactic dynamics.

Newtonian gravitational accelerations, which scale as $g_N \propto
MG/r^2$, transform under the above space-time scaling as
$g_N\rightarrow \l^{-2}g_N$, while kinematic accelerations ($\ddot x$)
transform as $g\rightarrow \l^{-1}g$. Newtonian dynamics, which
equates the two, is thus not scale invariant. To have such a symmetry
in the MOND limit we need $g$ to scale like $g_N^{1/2}$; or, with the
help of $a_0$, $g\propto (a_0 g_N)^{1/2}$. This, more primitive,
description of the MOND limit, is essentially the original formulation
of MOND for test-particle motion \cite{Milgrom83}. The formulation of
the 3rd tenet in terms of scale invariance\cite{milgrom09a}, is,
however, rather more elegant, precise, and general, and should be
preferred.

Beyond the above basic tenets, one wishes to construct a full-fledged
theory, and then extend it to the relativistic regime. Several
relativistic and NR MOND theories are known.  In the NR regime, we
have the suitably chosen nonlinear extension of the Poisson
equation\cite{bm84}, and a quasilinear MOND formulation
(QUMOND)\cite{milgrom10a}. These are classified as `modified gravity'
(MG) theories as they modify the field equations of the gravitational
field, but not the laws of motion. There are also `modified inertia'
(MI) formulations, which do the opposite. For the latter there isn't
yet a full fledged theory, but, nonetheless, much can be said about
their predictions of rotation curves\cite{milgrom94a,milgrom11}.  In
the relativistic regime we can mention as examples, TeVeS\cite{bek04},
and its predecessors, MOND adaptations of Einstein aether
theories\cite{zlosnik07}, bimetric MOND (BIMOND)\cite{milgrom09}, one
based on a polarizable medium\cite{blt09}, and non-local, single
metric formulations\cite{deffayet11}.  Much work has also been done
over the years towards devising observational tests of MOND and on
comparing MOND predictions with observations.  Extensive recent
reviews of these aspects of the MOND paradigm
exist\cite{Sanders02,Famaey12}.

Quantum theory and relativity were not mere changes in form of the
equations of classical dynamics. They each brought about totally new
concepts to underlie dynamics. Likewise, there are reasons to presume
that what we know about MOND today is only the tip of an
iceberg\cite{milgrom09a}. One hint that this might be so is the
`coincidence'\cite{Milgrom83} $\bar a_0\equiv 2\pi a_0\approx
c\,H_0\approx c^2(\Lambda/3)^{1/2}$, where $H_0$ is the Hubble
constant and $\Lambda$ is the cosmological constant (CC).  This, and
some aspects of symmetry, may point to a deep connection between
dynamics within local systems, such as galactic systems, and the state
of the universe at large. This connection, if firmly established,
could be the most far-reaching implication of MOND.\footnote{Such a
  cosmological connection might imply that various aspects of the
  MOND-standard-dynamics interplay may depend on cosmic time.}

The MOND theories mentioned above depart from pre-MOND dynamics in
that they introduce $a_0$, add new degrees of freedom, and modify the
underlying action, but they do not deeply depart in spirit from their
predecessor. Perhaps one of them will turn out to be an effective
theory that captures the essence of the future deeper MOND
theory.\footnote{The appearance in all these theories of an
  interpolating function is an indication that they can only be
  effective theories.}

In addition, there are several ideas and suggestions for a
`microscopic' basis of MOND.  One such class of
suggestions\cite{milgrom99} propounds that the MOND departure from
Newtonian dynamics arises from the fact that we do not live in a
globally Minkowski space time, but in one governed by a CC. The recent
discovery of a CC-like entity that affects cosmology points to the
possibility that the universe is entering an era where it will be
increasingly dominated by the CC. Such
an eventual universe, a de Sitter universe, is characterized by a
radius $R_\Lambda\equiv (\Lambda/3)^{-1/2}$. This fact might show
itself in local dynamics in the form of the acceleration
constant\cite{milgrom99} $\bar a_0=a_\Lambda\equiv c^2/R_\Lambda$, as
indeed appears in the above-mentioned `coincidence'\footnote{This
  appearance of $\Lambda$ is different from the standard appearance of
  $R_\Lambda$ in correcting the large distance behavior of (Newtonian)
  gravity introducing a linear, repulsive force.}. For example, in a
de Sitter universe, the quantum vacuum is modified from the
Minkowskian one in such a way that the Unruh radiation seen by an
accelerated observer depends both on $a_\Lambda$ and on the observer's
acceleration $a$ (assumed constant) in a way that may point to a MOND
inertia law\cite{milgrom99}.

There have been several recent proposals\cite{pikhitsa10,klinkhamer12}
for obtaining MOND dynamics by adding the element of the de Sitter
background\cite{milgrom99} to the idea of entropic-origin-of-dynamics
suggested by Verlinde\cite{jacobson95,padma10,verlinde11}.  In the
above schemes, the $a_0$ that appears in galactic-scale dynamics is
indeed related to the cosmological constant as $a_0\sim
a_\Lambda$\footnote{Note that $\sim$\ is used as a similarity sign
  indicating that two numbers are of the same order of magnitude,
  while $\approx$ means two numbers are similar within a factor of a
  few.}. In a different approach\cite{Pazy12}, also hinging on
entropic gravity, $a_0$ emerges as the Fermi energy on an holographic
screen.

In the MOND-through-a-polarizable-medium theory\cite{blt09}, in MOND
adaptations of Einstein-aether theories\cite{zlosnik07}, and in
BIMOND\cite{milgrom09}, the relation $a_0\sim a_\Lambda$ appears
differently: In these theories a length parameter appears in two
places in the action, one that ends up defining the CC, the other
defining $a_0$. If one insists, for economy or naturalness, to use the
same quantity in both roles, the above relation holds. However, there
is no compelling mechanism for forcing this economy.

In light of the uncertainty in what the final space-time-matter theory
will look like, it is reassuring that one can predict\cite{Milgrom08a,milgrom13}
a number of general dynamics laws, which compare excellently with
observations\cite{Sanders02,Famaey12} and are based almost entirely on
the basic tenets. These laws are important to recognize and understand
for several reasons: (1) They constitute the core predictions of MOND,
obeyed by all MOND theories, present and future, that embody the basic
tenets.  (2) They focus attention on well defined, easy-to-grasp
predictions of MOND. (3) They involve only limited information about
the systems (such as global attributes) and hence are easier to test
on large samples. (4) They constitute a list of `things to show' in
the competing paradigm of Newtonian-dynamics--plus-DM.
Succinctly formulated, these laws are:
\vspace{-6mm}
\begin{enumerate}[(i)]
\item{\label{asym}} Speeds along an orbit around any isolated mass,
  $M$, become asymptotically independent of the size-scale of the orbit. For
  circular orbits, $V(r\rightarrow\infty)\rightarrow V_{\infty}(M)$.
\item{\label{btfr}}

$V_{\infty}(M)=(M\,G\,a_0)^{1/4}$.
\item{\label{bound}}
In disc galaxies transition from `baryon dominance' to `DM
  dominance' occurs around the radius where $V^2(r)/r=a_0$.

\item{\label{max}}
The acceleration attributed to `DM' can never much exceed $a_0$.

\item{\label{centralsig}} The central surface density of `dark
  halos' is $\approx \Sigma_M\equiv a_0/ 2\pi G$.

\item{\label{siglimit}}
Quasi-isothermal (baryonic) systems have mean surface densities
$\bar\Sigma\lesssim \Sigma_M$.

\item{\label{msigma}} Quasi-isothermal or deep-MOND systems have a
  velocity dispersion $\sigma\approx (M\,G\,a_0)^{1/4}$.

\item{\label{efe}}
The external field in which a system is falling affects its intrinsic dynamics.

\item{\label{disc}} Disc galaxies behave as if they have both disc and
  spherical `DM' components.

\item{\label{stability}}
  MOND endows self gravitating systems with an increased, but limited
  stability.

\end{enumerate}
\vspace{-2mm} In addition to these laws, it is also reassuring that
even the more elaborate prediction of rotation curves in MOND follows
almost entirely from the basic premises alone\cite{milgrom13}.  The
accuracy of MOND's prediction of full rotation curves of disc galaxies
is particularly impressive, as shown by the analysis of over a hundred
galaxies\cite{Famaey12}.  Recently, MOND has also passed an acute test
in accounting for the dynamics of elliptical galaxies\cite{Milgrom12}.

\section{MOND vs dark matter}
\label{kpmsec:MONDdm}

We have seen that the SMoC faces many difficulties in accounting for
various aspects of data on galaxies (Sec.~\ref{kpmsec:DDGT} --
\ref{kpmsec:additionaltests}). But perhaps the most cogent argument
against the SMoC is that MOND
works\cite{Milgrom08a,Milgrom08b,Sanders09}
(e.g. Fig.~\ref{kpmfig:mda}). This is not because MOND is a
competitor, but because the many tight regularities and correlations
between baryon and `DM' properties, predicted by MOND, and confirmed
in the data, are quite contrary to what is expected in the DM
paradigm.  In the DM paradigm, a present-day system is the haphazard
end result of a complex history that affects baryons and DM very
differently. The observed tight correlations argue against
this\cite{Disney08}.

It is important to emphasize that in MOND, the dynamics of a
self-gravitating system is strictly predicted once its baryon
distribution is given. But in the SMoC it is not possible to predict
the dynamics of an individual primordial (type~A) galaxy just from its
observed baryons; at best one can try to match a DM halo from an
infinite family that will fit the observed dynamics. For example,
given the baryon distribution of a disc galaxy, the MOND rotation
curve is a prediction, while the DM curve is but a fit with much
freedom allowed.
In short, unlike for MOND, the SMoC cannot predict the relations
 between baryons and DM in any individual system because these depend
 on the unknowable history of the system.
In the rare case where the SMoC does make a prediction (tidal dwarfs,
i.e. type~B galaxies), because their formation histories are well
understood, it fails\cite{bournaud07}.

Angus\cite{Angus11} and others have demonstrated that MOND-based
cosmological models are able to account for many observations such as
the CMB and the Bullet cluster.

\section{MOND and the dwarfs}
\label{kpmsec:MONDdwarfs}
In MOND the dual dwarf galaxy theorem predicts no systematic dynamical
differences between primordial and tidal dwarf
galaxies\cite{Gentile07,mil07}.  For example, both rotational types
A~and~B of dwarfs are predicted to fall on the same BTFR by MOND
law~\ref{btfr} in Sect.~\ref{kpmsec:MOND}, irrespective of their
different formation paths.  MOND is fully consistent with the
existence of the VPOS around the Milky Way
(Sec.~\ref{kpmsec:phasespace}) as being an ancient tidal structure in
which the Galaxy's satellites formed as tidal dwarf galaxies and
globular clusters. Tidal dwarf galaxies form naturally in MONDian
galaxy interactions\cite{Tiret07,Combes10}. The internal constitution
of the Galaxy's dwarf-spheroidal satellites is well explained with
MOND\cite{McGaugh10,Hernandez10}. No such consistent explanation can
be achieved within the SMoC, because the high measured Newtonian
mass-to-light ratios of the satellites are completely in contradiction
with them being ancient tidal dwarf galaxies. This, however, is the
only physically plausible explanation for the existence of the
VPOS\cite{Pawlowski12a, Fouquet12} with counter-orbiting
satellites\cite{Pawlowski11}.

\section{Conclusions}
\label{kpmsec:concs}

The SMoC is based on five postulates (Sec.~\ref{kpmsec:introd}); it is
reported as having appreciable success in describing the large scale
structure and the CMB. More importantly, it has a long history of
grave failures\footnote{Compare to the standard model of particle
  physics which is known to be incomplete but since decades it has
  consistently been found to be in excellent agreement with the
  data.}. The failures are not really surprising since the first and
most fundamental postulate is equivalent to an extrapolation of the
empirically established gravitational law by Newton and Einstein to
scales many orders of magnitude below the Solar system acceleration
scale.  Instead, the data correlations unequivocally show dynamics at
accelerations smaller than $a_0\approx 1.2\times10^{-10}$m/s$^2$ to
not be Einsteinian/Newtonian.  Given the nature of the failures,
dynamically significant cold or warm DM particles cannot {be the
  explanation of the mass discrepancies}.  {Observations strongly
  suggest that a complete theory of dynamics has to depart from
  standard dynamics below some critical acceleration that is a
  constant of nature.} The extraordinary success of MOND in accounting
for the observational data on galactic scales and its properties
(Sec.~\ref{kpmsec:MOND}--\ref{kpmsec:MONDdwarfs}) suggest that the
expected underlying theory will contain a deep connection between the
dynamics within local systems and the state of the universe at large.
{Understanding} the deeper physical meaning of MOND remains a
challenging {aim.  It involves the realistic likelihood that a major
  new insight into gravitation will emerge, which would have
  signiﬁcant implications for our understanding of space, time and
  matter.}

%

\section{References}
\bibliographystyle{ws-ijmpd}
\bibliography{KroPaMil_refs}

\end{document}